\documentclass[lettersize,journal]{IEEEtran}
\usepackage{amsmath,amsfonts}
\usepackage{algorithmic}
\usepackage{algorithm}
\usepackage{array}
\usepackage[caption=false,font=normalsize,labelfont=sf,textfont=sf]{subfig}
\usepackage{textcomp}
\usepackage{stfloats}
\usepackage{url}
\usepackage{verbatim}
\usepackage{graphicx}
\usepackage{cite}
\usepackage{multirow}
\usepackage{titlesec}
\usepackage{xcolor}
\usepackage{fontawesome}

\titlespacing{\section}{0pt}{*0}{*0}
\titlespacing{\subsection}{0pt}{*0}{*0}
\titlespacing{\subsubsection}{0pt}{*0}{*0}

\begin{document}

\title{Hierarchy-Aware and Channel-Adaptive Semantic Communication for Bandwidth-Limited Data Fusion\\}

\author{\IEEEauthorblockN{Lei Guo, Wei Chen, \IEEEmembership{Senior Member, IEEE}, Yuxuan Sun, \IEEEmembership{Member, IEEE}, Bo Ai, \IEEEmembership{Fellow, IEEE}, \\
Nikolaos Pappas, \IEEEmembership{Senior Member, IEEE}, Tony Quek, \IEEEmembership{Fellow, IEEE}}
\thanks{
Lei Guo, Wei Chen and Bo Ai are with the State Key Laboratory of Advanced Rail Autonomous Operation, Beijing Jiaotong University, China and the School of Electronic and Information Engineering, Beijing Jiaotong University, Beijing, China. Yuxuan Sun is with the School of Electronic and Information Engineering, Beijing Jiaotong University, Beijing, China. (e-mail:leiguo@bjtu.edu.cn; weich@bjtu.edu.cn; yxsun@bjtu.edu.cn; boai@bjtu.edu.cn). Corresponding author: Wei Chen, Yuxuan Sun.

Nikolaos Pappas is with the Department of Computer and Information Science, Linköping University, Linköping, Sweden (e-mail:nikolaos.pappas@liu.se).

Tony Quek is with the Information Systems Technology and Design, Singapore University of Technology and Design, Singapore (e-mail:tonyquek@sutd.edu.sg).

This work was supported in part by the Beijing Natural Science Foundation under grant L222044; the Natural Science Foundation of China (U2468201, W2421083, 62221001, 62301024) and the Talent Fund of Beijing Jiaotong University under grant 2023XKRC030.
}
}


\maketitle

\begin{abstract}
Obtaining high-resolution hyperspectral images (HR-HSI) is costly and data-intensive, making it necessary to fuse low-resolution hyperspectral images (LR-HSI) with high-resolution RGB images (HR-RGB) for practical applications. However, traditional fusion techniques, which integrate detailed information into the reconstruction, significantly increase bandwidth consumption compared to directly transmitting raw data. To overcome these challenges, we propose a hierarchy-aware and channel-adaptive semantic communication approach for bandwidth-limited data fusion. A hierarchical correlation module is proposed to preserve both the overall structural information and the details of the image required for super-resolution. This module efficiently combines deep semantic and shallow features from LR-HSI and HR-RGB. To further reduce bandwidth usage while preserving reconstruction quality, a channel-adaptive attention mechanism based on Transformer is proposed to dynamically integrate and transmit the deep and shallow features, enabling efficient data transmission and high-quality HR-HSI reconstruction. Experimental results on the CAVE and Washington DC Mall datasets demonstrate that our method outperforms single-source transmission, achieving up to a 2 dB improvement in peak signal-to-noise ratio (PSNR). Additionally, it reduces bandwidth consumption by two-thirds, confirming its effectiveness in bandwidth-constrained environments for HR-HSI reconstruction tasks.

%Semantic communication is the process of conveying useful information in the semantic domain. In most existing works, semantic features are designed to be learned and transmitted for single source data. However, these approaches are not suitable for practical communication systems that involve data from more than one source. Most multiple source data has more correlation compared to single source data. When transmitting different data sources, redundancy of the data can be removed to reduce bandwidth consumption. At the same time, different data sources can complement each other to improve task performance. In this paper, we propose a novel data fusion semantic communication system for different data sources to address this challenge. The proposed system is designed to remove redundancy across different data sources and transmit the fused semantic feature at the transmitter. The fused feature is then reconstructed into the data at the receiver. Experimental results demonstrate that the data fusion semantic communication achieves better performance than single source communication that only transmits the features of a single source under the same bandwidth. And data fusion semantic communication has lower performance degradation at lower bandwidths.
\end{abstract}

\begin{IEEEkeywords}
Semantic communication, hierarchy-aware data fusion, deep learning, HR-HSI reconstruction.
\end{IEEEkeywords}

\section{Introduction}
\IEEEPARstart{H}{yperspectral} super-resolution plays a critical role in enhancing the quality and detail of hyperspectral images (HSI) for applications like remote sensing. By combining high-resolution RGB (HR-RGB) images, high-resolution HSI (HR-HSI) can be reconstructed from low-resolution HSI (LR-HSI). This fusion takes advantage of the spatial detail from HR-RGB and the rich spectral information from LR-HSI, making it essential for acquiring high-quality data more efficiently. Advanced techniques, such as convolutional neural networks (CNNs), Transformers, and graph convolutional networks (GCNs), have been employed to improve feature extraction and the fusion process \cite{9895316, 10264151, 10433704}. For example, multiple convolutional kernels in a CNN are used to extract and fuse features for enhanced super-resolution \cite{9895316}. Additionally, the Transformer-based module enables the establishment of long-range dependencies in the spectral dimension during fusion \cite{10264151}, while the proposed GCN-based fusion method in \cite{10433704} extends this capability to capture long-range dependencies in both spatial and spectral dimensions. However, these methods neglect the increased data dimensionality of fusion, which may lead to higher bandwidth requirements than raw data. Shallow features capture details of the image, while deep features extract the overall structural information of the image \cite{8953615}. This issue is particularly challenging for HSI super-resolution tasks, which require both deep and shallow features, posing further challenges in bandwidth-limited environments.

In bandwidth-limited scenarios, traditional communication methods become inefficient as the complexity and size of fused data overwhelm available bandwidth, limiting real-time requirements in satellite communications and other time-sensitive applications \cite{9428096}. Semantic communication (SC) has emerged as a paradigm shift to address these issues \cite{ICASGoalOriented}. The semantic features of data are transmitted to reduce bandwidth requirements significantly \cite{10328187,10772628}. SC enhances data fusion processes and promotes resource efficiency, improving the scalability of intelligent systems by focusing on the context and importance of data \cite{LuoPIMRC24, 10454584, LuoWiOpt24}.

% \cite{JankowskiMikolaj0Wireless, 10118965, 9953316}
SC has been highly effective in single-modal data transmission, and recent research has expanded its application in multi-modal fusion tasks. For example, cross-modal coding for semantic-level fusion is used in \cite{10233481} for different task requirements of users. To optimize transmission efficiency, dynamic schemes adjust transmitted features based on channel conditions across multiple modalities in \cite{Guo2025videoQA}. Further advancements include a distributed system for audio-visual parsing \cite{10462495} and channel-level fusion for multi-modal data integration \cite{9921202}. Additionally, the MU-DeepSC system integrates image and text for visual question answering \cite{9653664}, while a novel layer-wise Transformer architecture improves data fusion by connecting each layer of the encoder and decoder, supporting multi-user environments \cite{9830752}. However, these approaches have not fully addressed the unique challenges of hyperspectral super-resolution tasks based on fusion, where the fused information exceeds the size of the original raw data, leading to increased transmission demands.

This letter presents a semantic communication approach for hierarchy-aware and channel-adaptive data fusion. It adaptively combines deep features with shallow features to improve both communication efficiency and reduces data redundancy. Our main contributions can be summarized as:
\begin{itemize}
%\item[$\bullet$] We propose a novel semantic communication system that employs hierarchy-aware data fusion to adaptively integrate data from diverse sources, significantly improving the performance of super-resolution task.

%\item[$\bullet$] Our system introduces a Hierarchically Interwoven Transformer (HIT) fusion mechanism, which utilizes Transformer to effectively combine deep semantic features with shallow detailed features. This fusion enhances feature selection, ensuring the transmission of critical semantic information and essential super-resolution details, while reducing bandwidth usage and data redundancy.

\item[$\bullet$] To address the bandwidth challenges in hyperspectral super-resolution tasks based on LR-HSI and HR-RGB fusion, we propose a semantic communication based data fusion mechanism which is hierarchy-aware and channel-adaptive. This approach improves the transmission efficiency of fused information without increasing bandwidth usage, enabling high-quality HR-HSI reconstruction.

\item[$\bullet$] We propose a channel-adaptive attention mechanism based on Transformer, which adaptively selects the features across diverse layers and integrates deep features with shallow features to enhance data fusion. Experimental results show that our method outperforms single-modal system. Furthermore, compared to methods without the module, our approach reduces bandwidth consumption by two-thirds with less than 1 dB loss in reconstruction quality.

\end{itemize}

\section{System Model}

\begin{figure}[!t]
\centering
\includegraphics[width=\linewidth]{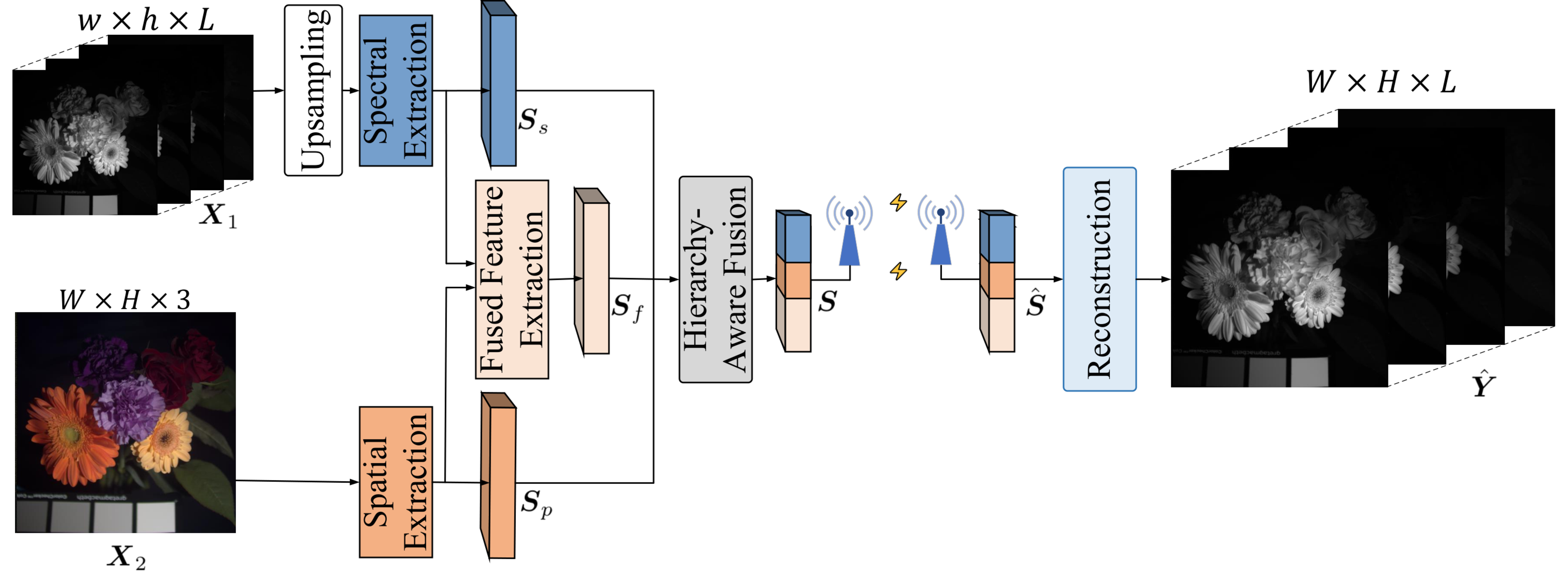}
\caption{The proposed hierarchy-aware and channel-adaptive semantic communication architecture for bandwidth-limited data fusion.}
\label{fig:figure1}
\end{figure}
%[width=2.5in]

%\begin{figure}
%\centering
%\includegraphics[width=\linewidth]{framework-modal.pdf}
%\caption{The proposed selective data fusion semantic communication system.}
%\label{fig:figure1}
%\end{figure}

The proposed hierarchy-aware and channel-adaptive semantic communication architecture for bandwidth-limited data fusion is illustrated in Fig. \ref{fig:figure1}. In this architecture, deep features capture the overall structural information necessary for the image reconstruction, while shallow features provide essential details crucial for super-resolution. To enhance communication efficiency, the architecture adaptively integrates fused features with shallow features from diverse data sources, effectively obtaining both structural and detailed information for high-quality image reconstruction. Our system employs LR-HSI and HR-RGB as the primary sources for data fusion. Additionally, we extract the spectral features from the LR-HSI, valued for their spectral data, and the spatial features from the HR-RGB, recognized for their spatial resolution. These features are adaptively integrated at the transmitter, forming the transmitted semantic information. %\textcolor{blue}{Beyond HSIs, the proposed semantic communication framework offers broader applications. In autonomous driving, it enables real-time fusion of sensor data from LiDAR, cameras, and radar \cite{8957313}. Similarly, in medical imaging, the framework enhances the integration and efficient transmission of multimodal medical image data, supporting remote diagnostics \cite{10505722}.}

We represent the LR-HSI and HR-RGB as $\boldsymbol{X}_{1}\in \mathbb{R}^{w \times h \times L}$ and $\boldsymbol{X}_{2}\in \mathbb{R}^{W \times H \times 3}$, respectively, which are sampled at the same transmitter. $\boldsymbol{Y}\in \mathbb{R}^{W \times H \times L}$ is represented as the corresponding HR-HSI, which are reconstructed at the receiver. In this notation, $w \ll W$ and $h \ll H$ indicate smaller spatial dimensions of LR-HSI relative to HR-HSI, while $3 \ll L$ suggests a greater spectral depth in HSI, providing detailed spectral and spatial information. 

To better extract detailed features, the spatial and spectral feature extraction modules at the transmitter employ distinct modules for their respective domains. Spectral features capture channel-dependent information from LR-HSI, while spatial features emphasize high-resolution structural details from HR-RGB. The system includes a fused feature extraction module, a spectral feature extraction module, and a spatial feature extraction module, represented as $S^{e}_{f}(\cdot;\boldsymbol{\alpha}_{f})$, $S^{e}_{s}(\cdot;\boldsymbol{\theta}_{s})$ and $S^{e}_{p}(\cdot;\boldsymbol{\delta}_{p})$, respectively. The parameters $\boldsymbol{\alpha}_{f}$, $\boldsymbol{\theta}_{s}$ and $\boldsymbol{\delta}_{p}$ correspond to the trainable parameters of each feature extraction module. Each module is specialized to handle distinct data inputs. The spectral module $S^{e}_{s}(\cdot;\boldsymbol{\theta}_{s})$ extracts spectral details from LR-HSI, while the spatial module $S^{e}_{p}(\cdot;\boldsymbol{\delta}_{p})$ focuses on capturing high-resolution spatial details from HR-RGB. The fused feature extraction module $S^{e}_{f}(\cdot;\boldsymbol{\alpha}_{f})$ integrates these detailed features to extract the overall structural features for improved data reconstruction. To better capture the overall structural information of the image, the fused deep features, i.e., $\boldsymbol{S}_{f}$  are obtained by feeding the shallow detail features, i.e., $\boldsymbol{S}_{s}$ and $\boldsymbol{S}_{p}$, through several residual blocks, which apply attention mechanisms across both spectral and spatial dimensions. These feature extraction processes  can be written as: 
\begin{equation}
\begin{aligned}
\boldsymbol{S}_{s} &= S^{e}_{s}\left(\boldsymbol{X}_{1};\boldsymbol{\theta}_{s}\right),
\boldsymbol{S}_{p} = S^{e}_{p}\left(\boldsymbol{X}_{2};\boldsymbol{\delta}_{p}\right),\\
\boldsymbol{S}_{f} &= S^{e}_{f}\left(\boldsymbol{S}_{s}, \boldsymbol{S}_{p};\boldsymbol{\alpha}_{f}\right),\\
\end{aligned}
\end{equation}
where $\boldsymbol{S}_{s}, \boldsymbol{S}_{p}, \boldsymbol{S}_f \in \mathbb{R}^{w_{1} \times h_{1} \times l}$ represent the features of spectral, spatial and the fusion modules, respectively. Here, $w_{1} \times h_{1}$ denotes the size of the feature map, while $l$ denotes the number of channels of the feature maps, where $l/L \leq 1/2$ indicating a reduction in channel dimensionality during the feature extraction process. 

To enable more flexible feature fusion under bandwidth constraints, a hierarchy-aware data fusion module is introduced to adaptively integrate both fused deep features and shallow features. The objective is to enhance HR-HSI reconstruction performance without increasing bandwidth usage, ensuring the efficient transmission of critical semantic features. This approach allows for efficient utilization of the available bandwidth by adaptively transmitting these features that contribute significantly to the quality of the HR-HSI. The hierarchical fusion process can be represented as:
\begin{equation}
\boldsymbol{S} = F\left(\boldsymbol{S}_{f}, \boldsymbol{S}_{s}, \boldsymbol{S}_{p};\boldsymbol{\beta}\right),
\end{equation}
where $\boldsymbol{S} \in \mathbb{R}^{w_{1} \times h_{1} \times l}$ represents the transmitted features after the hierarchy-aware fusion, maintaining the same dimensions as the individual input features, thereby avoiding increase in bandwidth consumption typically associated with data fusion. The fusion process is managed by the hierarchy-aware fusion module, denoted by $F\left(\cdot;\boldsymbol{\beta}\right)$, where $\boldsymbol{\beta}$ represents the trainable parameters of the module.

We consider the additive white Gaussian noise (AWGN) channel with Gaussian noise power $\sigma^2$. At the receiver, the fused feature is received, denoted by $\hat{\boldsymbol{S}} = \boldsymbol{S} + \boldsymbol{N}_0$ where $\boldsymbol{N}_0$ represents the noise. The fusion decoder, which utilizes trainable parameters $\boldsymbol{\rho}$, is implemented through a trainable decoder network represented as $D(\cdot;\boldsymbol{\rho})$. The HR-HSI reconstructed from the adaptive fusion features can be expressed as:
\begin{equation}
\hat{\boldsymbol{Y}}=D\left(\hat{\boldsymbol{S}};\boldsymbol{\rho}\right),
\end{equation}
where $\hat{\boldsymbol{Y}} \in \mathbb{R}^{W \times H \times L}$ denotes the reconstructed HR-HSI, representing the spatial and spectral enhancement achieved through fusion. The primary objective of our proposed data fusion semantic communication system is to accurately reconstruct the HR-HSI from LR-HSI and HR-RGB. We employ the mean square error (MSE) as the loss function. MSE measures the discrepancy between the ground truth HR-HSI, $\boldsymbol{Y}$, and the reconstructed HR-HSI, $\hat{\boldsymbol{Y}}$, and is formulated as follows:
%The objective of the proposed data fusion semantic communication is to reconstruct the HR-HSI based on LR-HSI and HR-RGB. The proposed transceiver is task-oriented, where the HR-HSI is reconstructed directly at the receiver. The mean square error (MSE) is used as the loss function to measure the difference between the ground truth, $\boldsymbol{Y}$, and the reconstructed HR-HSI, $\hat{\boldsymbol{Y}}$, which can be formulated as:
\begin{equation}
\mathcal{L}_{\text{MSE}}(\boldsymbol{Y},\hat{\boldsymbol{Y}};\boldsymbol{\alpha}_{f}, \boldsymbol{\theta}_{f}, \boldsymbol{\delta}_{f}, \boldsymbol{\beta}, \boldsymbol{\rho}) = \left\|\boldsymbol{Y}-\hat{\boldsymbol{Y}}\right\|^{2}_{2},
\end{equation}
where $\left\|\boldsymbol{\cdot}\right\|_{2}$ is the $l_{2}$-norm. 
%\widehat{\boldsymbol{Y}}

\begin{figure}[!t]
\centering
\includegraphics[width=0.6\linewidth]{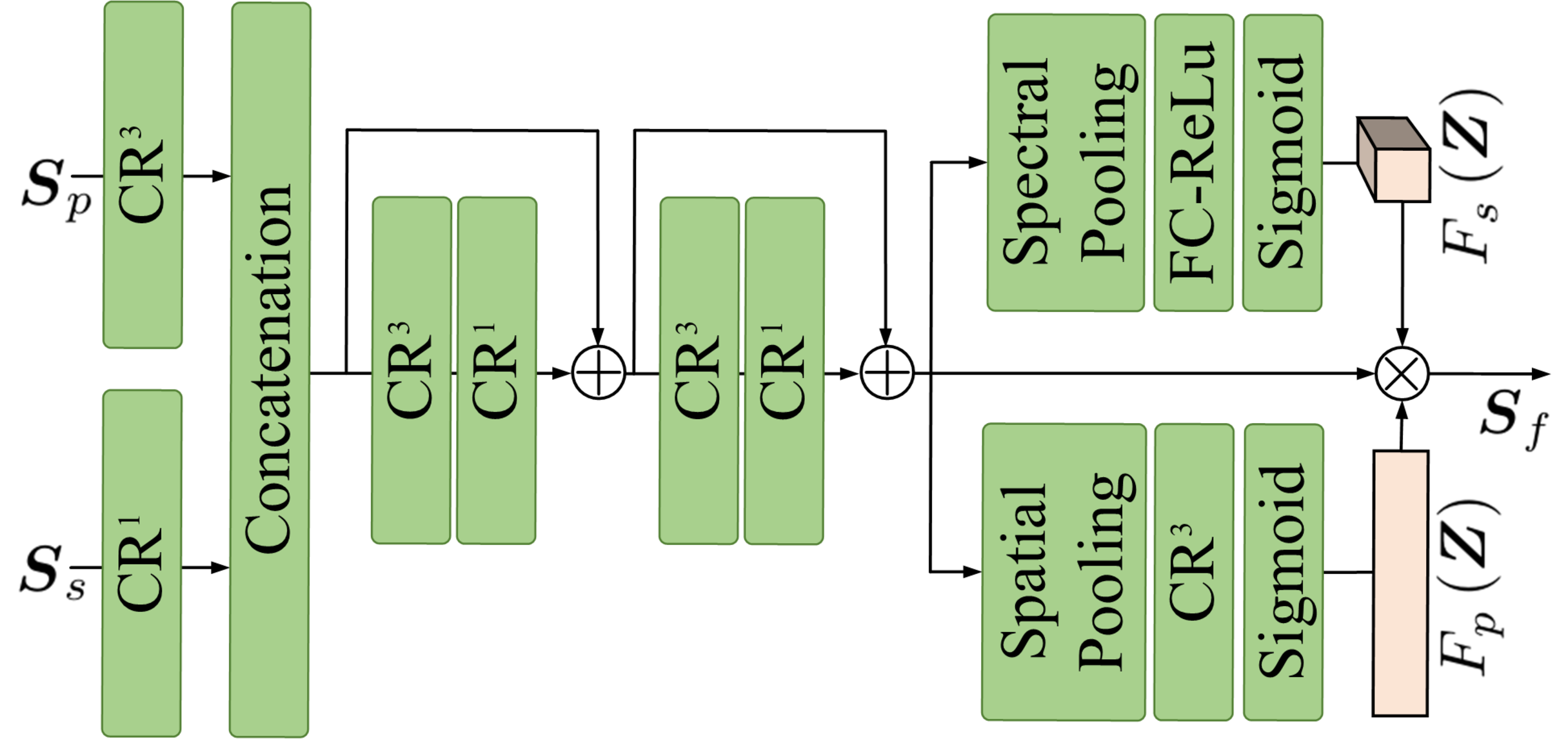}
\caption{The detailed structure of the proposed fused feature extraction module at the transmitter. $F_{{cr}^{*}}\left(\cdot\right)$  is represented as $\text{CR}^{*}$. $\bigoplus$ and $\bigotimes$ denote addition and multiplication of the same position elements, respectively.}
\label{fusion_encoder}
\end{figure}

%\begin{figure}
%\centering
%\includegraphics[width=0.8\linewidth]{fusion_FFE.pdf}
%\caption{The detailed structure of the proposed the fusion feature extraction module at the transmitter.}
%\label{fusion_encoder}
%\end{figure}

\section{The Details of the Hierarchy-Aware Data Fusion Semantic Communication}
In this section, we present our semantic communication framework that integrates data fusion of HR-RGB and LR-HSI. The framework employs a hierarchy-aware fusion structure to enhance communication efficiency, especially in bandwidth-constrained wireless environments. The fusion module within the framework adaptively integrates deep structural features with shallow detailed features.

\subsection{The Detailed Feature Extraction Modules}
At the transmitter, the LR-HSI is initially upsampled to align with the spatial dimensions of the HR-HSI. A shallower network architecture is utilized to effectively preserve detailed information in spatial and spectral feature extraction modules, where two convolution layers with rectified linear unit (ReLU) as activation functions are implemented. In the spatial extraction module, convolution with a kernel size of 3 is employed to capture adjacent spatial information. Meanwhile, convolution with a kernel size of 1 in the spectral feature extraction module is used to enhance the extraction of inter-spectral correlations. The spatial and spectral feature extraction modules are represented as follows:
\begin{equation}
\begin{aligned}
\boldsymbol{S}_{s} &= S^{e}_{s}\left(\boldsymbol{X}_{1};\boldsymbol{\theta}_{s}\right) = F_{{cr}^{1}}\left(F_{{cr}^{1}}\left(\boldsymbol{X}_{1}\right)\right),\\
\boldsymbol{S}_{p} &= S^{e}_{p}\left(\boldsymbol{X}_{2};\boldsymbol{\delta}_{p}\right) = F_{{cr}^{3}}\left(F_{{cr}^{3}}\left(\boldsymbol{X}_{2}\right)\right),
\end{aligned}
\end{equation}
where $F_{{cr}^{1}}\left(\cdot\right)$ and $F_{{cr}^{3}}\left(\cdot\right)$ denote the convolution layers with a kernel size of 1 and 3, respectively, both paired with ReLU activation functions.

\subsection{The Fused Feature Extraction Modules}
Compared to the spatial and spectral feature extraction modules, which primarily focus on extracting detailed information, the fused feature extraction module exhibits more complexity and depth, as shown in Fig \ref{fusion_encoder}. To fully leverage the spectral features of LR-HSI and the spatial characteristics of HR-RGB, we preprocess these features separately using $F_{{cr}^{*}}\left(\cdot\right)$ before concatenation. After concatenation and the initial feature fusion, we apply enhancement factors separately in the spatial and spectral domains to further strengthen these aspects. The process is mathematically represented as follows:
\begin{equation}
\begin{aligned}
&\boldsymbol{Z} = F_{f}\left( F_{f}\left( \left[F_{{cr}^{1}}\left(\boldsymbol{S}_{s}\right), F_{{cr}^{3}}\left(\boldsymbol{S}_{p}\right)\right] \right)\right),\\
&\boldsymbol{S}_{f} = \boldsymbol{Z} \times F_{s}\left( \boldsymbol{Z} \right) \times F_{p}\left( \boldsymbol{Z} \right),
\end{aligned}
\end{equation}
where $\boldsymbol{Z}$ denotes the initial fused feature and $\left[ \cdot \right]$ indicates the concatenation operation and $\times$ denotes element-wise multiplication. $F_{f}\left(\cdot\right)$ is a residual block, which can be expressed as $F_{f}\left(x\right) = x + F_{{cr}^{3}}\left(F_{{cr}^{1}}\left(x\right)\right)$. As illustrated in Fig. \ref{fusion_encoder}, $F_{s}\left( \cdot \right)$ and $F_{p}\left( \cdot \right)$ represent the spectral and spatial enhancement factor modules, respectively.

\subsection{Channel-Adaption in Hierarchy-Aware Fusion}
\begin{figure}[!t]
\centering
\includegraphics[width=0.6\linewidth]{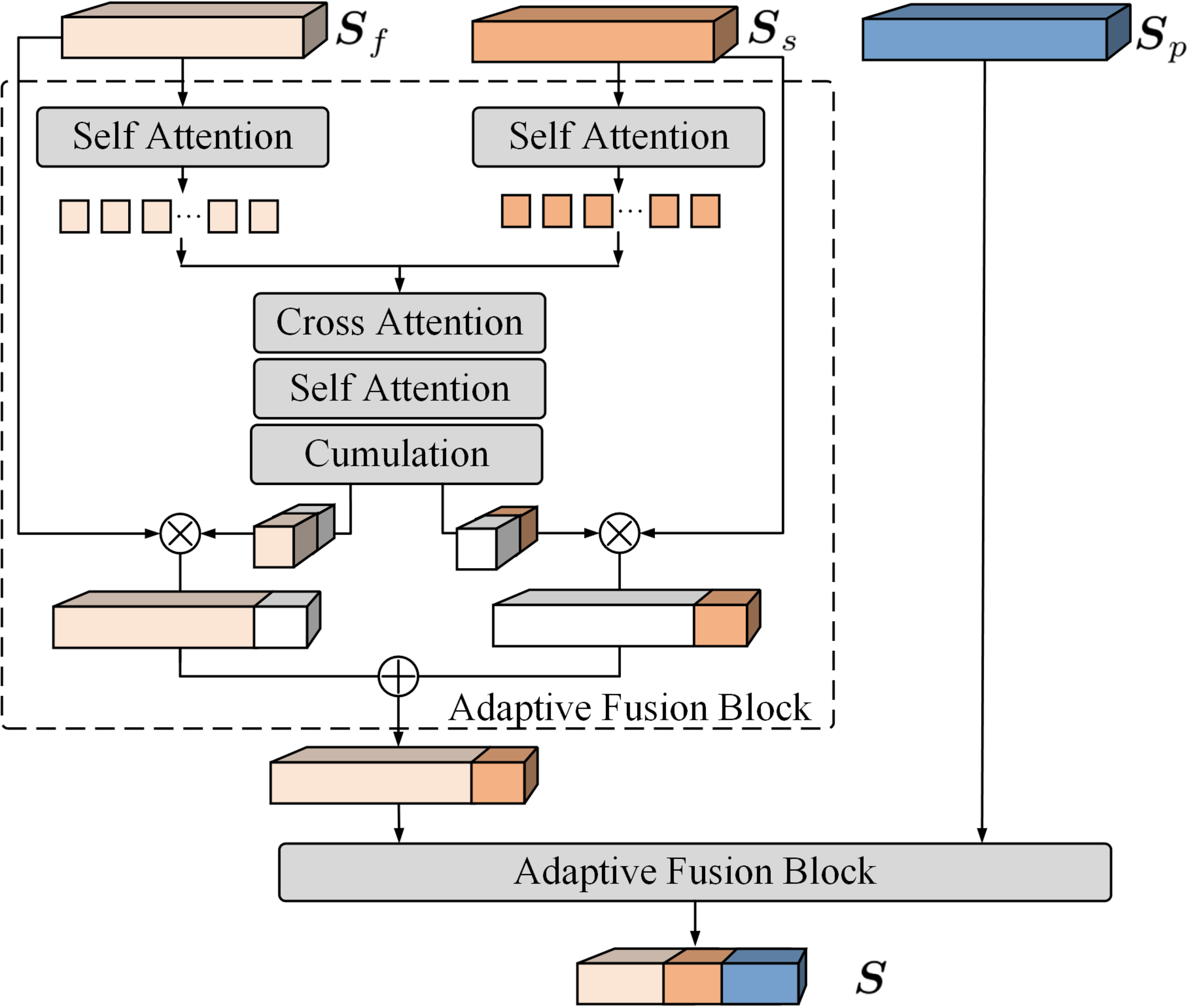}
\caption{The detailed structure of the proposed adapative fusion block in the hierarchy-aware fusion module.}
\label{fusion_SFE}
\end{figure}

%\begin{figure}
%\centering
%\includegraphics[width=0.8\linewidth]{fusion_SFE.pdf}
%\caption{The detailed structure of the proposed the selective fusion module at the transmitter.}
%\label{fusion_SFE}
%\end{figure}

The hierarchy-aware fusion module is a critical innovation in our semantic communication system, designed to adaptively integrate the overall structural and detailed features using advanced Transformers without additional bandwidth consumption, as shown in Fig. \ref{fusion_SFE}. Our methodology employs a hierarchy-aware fusion module that adaptively combines deep and shallow features to simultaneously transmit structural and detailed features under bandwidth constraints. The deep features extract the overall structural context, while the shallow features preserve spectral and spatial information details.

This module includes two adaptive fusion blocks that utilize self-attention (SA) \cite{2017Attention} and cross-attention (CA) \cite{jaegle2021perceiver
} mechanisms to integrate critical features. A Transformer-based attention module computes the dynamic guidance vector $\boldsymbol{N}_{(f, s)}, \boldsymbol{N}_{((f,p),s)} \in \mathbb{R}^{w_1 \times h_1 \times l}$ via an SA-CA-SA sequence with softmax function. This vector is used to generate a mask feature $\boldsymbol{M}_{(f,p)}, \boldsymbol{M}_{((f,p),s)} \in \mathbb{R}^{w_1 \times h_1 \times l}$ via a cumulative process, where the subscript $(f, p)$ indicates the integration of $\boldsymbol{S}_f$ and $\boldsymbol{S}_p$. The subscript $\left(\left(f, p\right), s\right)$ indicates the further integration with $\boldsymbol{S}_s$ after the integration of $\boldsymbol{S}_f$ and $\boldsymbol{S}_p$. This mask feature controls the adaptive fusion of deep and shallow features, mathematically expressed as: 
\begin{equation} 
\boldsymbol{M}_{(f,p)}^{(i)} = \begin{cases} \boldsymbol{N}_{(f,p)}^{(i)} & \text{for } i=1, \\
\boldsymbol{M}_{(f,p)}^{(i-1)} + \boldsymbol{N}_{(f,p)}^{(i)} & \text{for } i=2,\ldots,l, 
\end{cases} 
\end{equation} 
where $\boldsymbol{M}_{(f,p)}^{(i)}$ denotes the values in the $i$-th channel of the feature. Similarly, $\boldsymbol{M}_{((f,p),s)}$ is also calculated in this way. The adaptive fusion process combines these features $\boldsymbol{S}_f$, $\boldsymbol{S}_s$ and $\boldsymbol{S}_p$ as follows:
\begin{equation} 
\begin{aligned} 
&\boldsymbol{S}_{(f,p)} = \boldsymbol{S}_f \times \boldsymbol{M}_{(f,p)} + \boldsymbol{S}_p \times \left(\boldsymbol{1} - \boldsymbol{M}_{(f,p)}\right), \\ 
&\boldsymbol{S} = \boldsymbol{S}_{(f,p)} \times \left(\boldsymbol{1} - \boldsymbol{M}_{((f,p),s)}\right) + \boldsymbol{S}_s \times \boldsymbol{M}_{((f,p),s)},
\end{aligned} 
\end{equation} 
where $\boldsymbol{1} \in \mathbb{R}^{w_{1} \times h_{1} \times l}$ is an array where each element equals one. The dynamic fusion process adaptively integrates feature maps to preserve critical structural and detailed features, achieving a two-thirds reduction in transmitted data volume with marginal performance degradation, ensuring transmission efficiency under bandwidth constraints.

\subsection{The Reconstruction Module}
\begin{figure}[!t]
\centering
\includegraphics[width=0.7\linewidth]{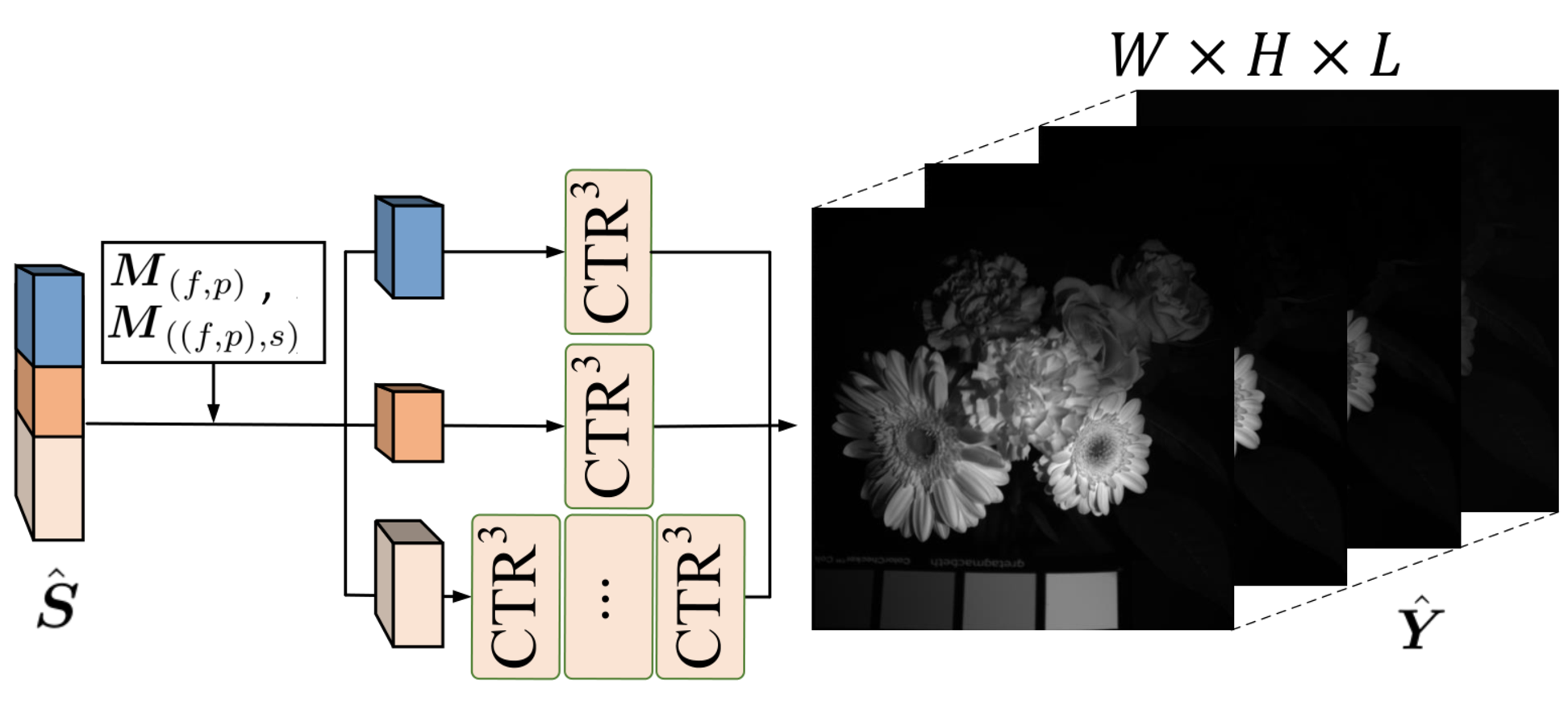}
\caption{The structure of the fusion decoder. $F_{{ctr}^{3}}\left(\cdot\right)$  is represented as $\text{CTR}^{3}$}
\label{fusion_decoder}
\end{figure}

At the receiver end, the fusion decoder is designed with reconstructing HR-HSI from the received noisy fused features. The fusion masks are assumed to be known at the receiver to enhance the reconstruction of detailed information. These masks occupy negligible bandwidth compared to the bandwidth required for transmitting the reconstructed features. The fusion decoder utilizes these masks to adaptively extract detailed information, which significantly enhances the quality of the reconstructed images during the decoding. The decoder incorporates a series of convolutional transpose layers, each equipped with ReLU activation, as shown in Fig. \ref{fusion_decoder}. These layers are designed to expand the dimensions of the received features by integrating residuals from the detailed features. This approach ensures a more detailed and accurate reconstruction of the HR-HSI, which can be expressed as:
\begin{equation}
\begin{aligned}
\hat{\boldsymbol{Y}} & =D\left(\hat{\boldsymbol{S}};\boldsymbol{\rho}\right)=F_{ctr^{3}}\left( F_{ctr^{3}}\left(\hat{\boldsymbol{S}}\right)\right) \\
& \!+\!F_{ctr^{3}}\left(\!\hat{\boldsymbol{S}}\!\times \!\boldsymbol{M}_{\left(f,p\right)}\! \right)\!+ \!F_{ctr^{3}}\left(\!\hat{\boldsymbol{S}}\!\times\!\left(\!\boldsymbol{1}\!-\!\boldsymbol{M}_{\left(\left(f,p\right),s\right)}\!\right)\!\right),
\end{aligned}
\end{equation}
where $F_{{ctr}^{3}}\left(\cdot\right)$ denotes the convolution transpose layers with the ReLU activation function.

\section{Results}

We evaluate single-source \cite{9438648} and data fusion semantic communication strategies for reconstructing HR-HSI using the CAVE \cite{5439932} and Washington DC Mall datasets \cite{DCMallDataset}. The CAVE contains 512 $\times$ 512 pixel with 31 spectral bands, while the Washington DC Mall dataset consists of 128 $\times$ 128 pixel with 191 spectral bands sampled by 1280 $\times$ 307 pixel HSIs. In our experiments, the original HSIs are used as the HR-HSI $\boldsymbol{Y}$ and are down-sampled to generate LR-HSI $\boldsymbol{X}_{1}$. Specifically, for both the CAVE and DC datasets, the $\boldsymbol{X}_{1}$ is down-sampled to $32 \times 32$. The experiments are conducted in the AWGN channel, Rayleigh channels with MSE equalization and MIMO channels with singular value decomposition (SVD). Reconstruction quality is primarily evaluated using the peak signal-to-noise ratio (PSNR) and the structural similarity index measure (SSIM) \cite{1284395}. Computational simulations were performed on Intel Xeon Silver 4110 CPU @2.10GHz and an NVIDIA A40 GPU.

\begin{figure}
    \centering
    \captionsetup[subfloat]{font=scriptsize,labelfont=scriptsize,textfont=scriptsize} % Ensure subfigure captions match main text settings
    \subfloat[Comparison on the CAVE dataset.]{\label{results_1} 
        \includegraphics[width=0.23\textwidth]{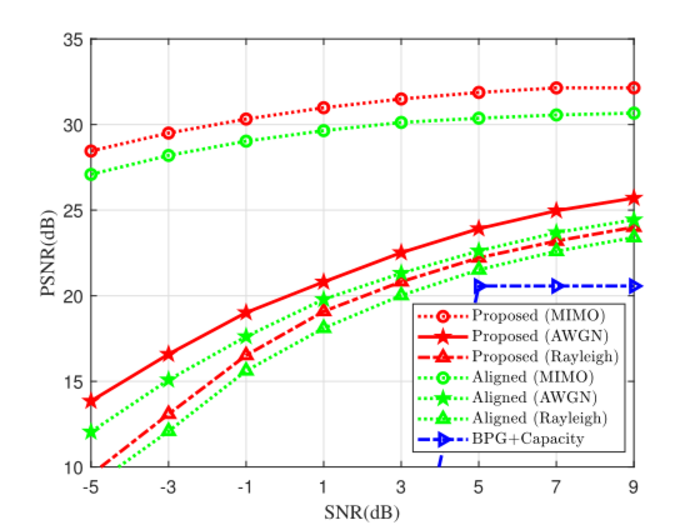}
    }
    \hfill
    \subfloat[Comparison on the DC dataset.]{\label{results_2} 
        \includegraphics[width=0.23\textwidth]{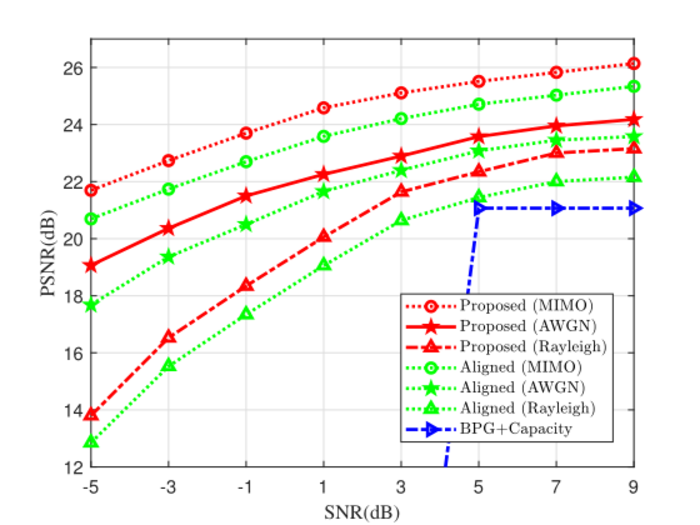}
    }
    \caption{Performance comparisons on the CAVE and DC dataset.}
    \label{results_1}
\end{figure}

\begin{figure}
    \centering
    \captionsetup[subfloat]{font=scriptsize,labelfont=scriptsize,textfont=scriptsize} % Ensure subfigure captions match main text settings
    \subfloat[PSNR performance.]{ 
        \includegraphics[width=0.23\textwidth]{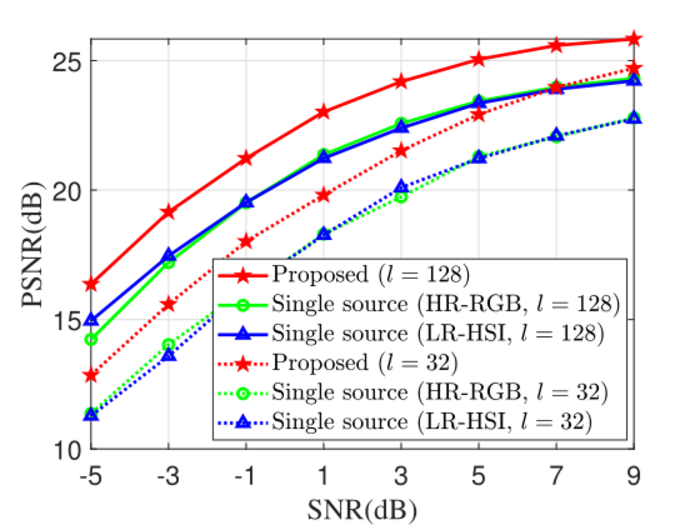}
    }
    \hfill
    \subfloat[SSIM performance.]{
        \includegraphics[width=0.23\textwidth]{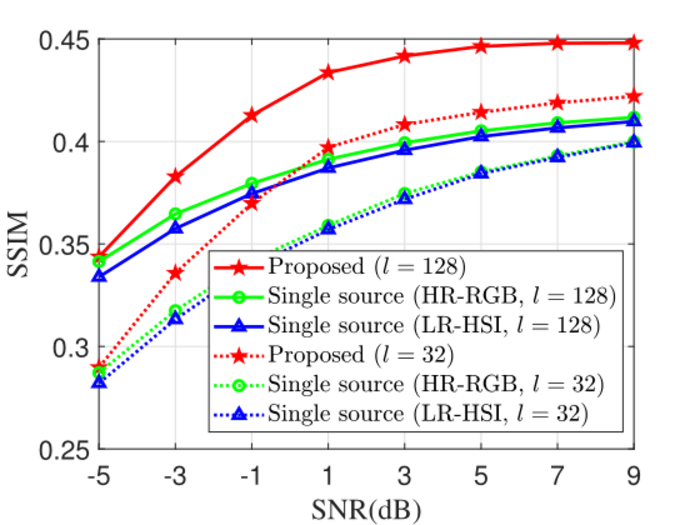}
    }
    \caption{Performance comparisons between single-source and data fusion semantic communication methods for different values of $l$..}
    \label{results_2}
\end{figure}

\begin{figure}[t]
    \centering
    \captionsetup[subfloat]{font=scriptsize,labelfont=scriptsize,textfont=scriptsize} % Ensure subfigure captions match main text settings
    \subfloat[Visualized result on CAVE.]{\label{results_cave} 
        \includegraphics[width=0.23\textwidth]{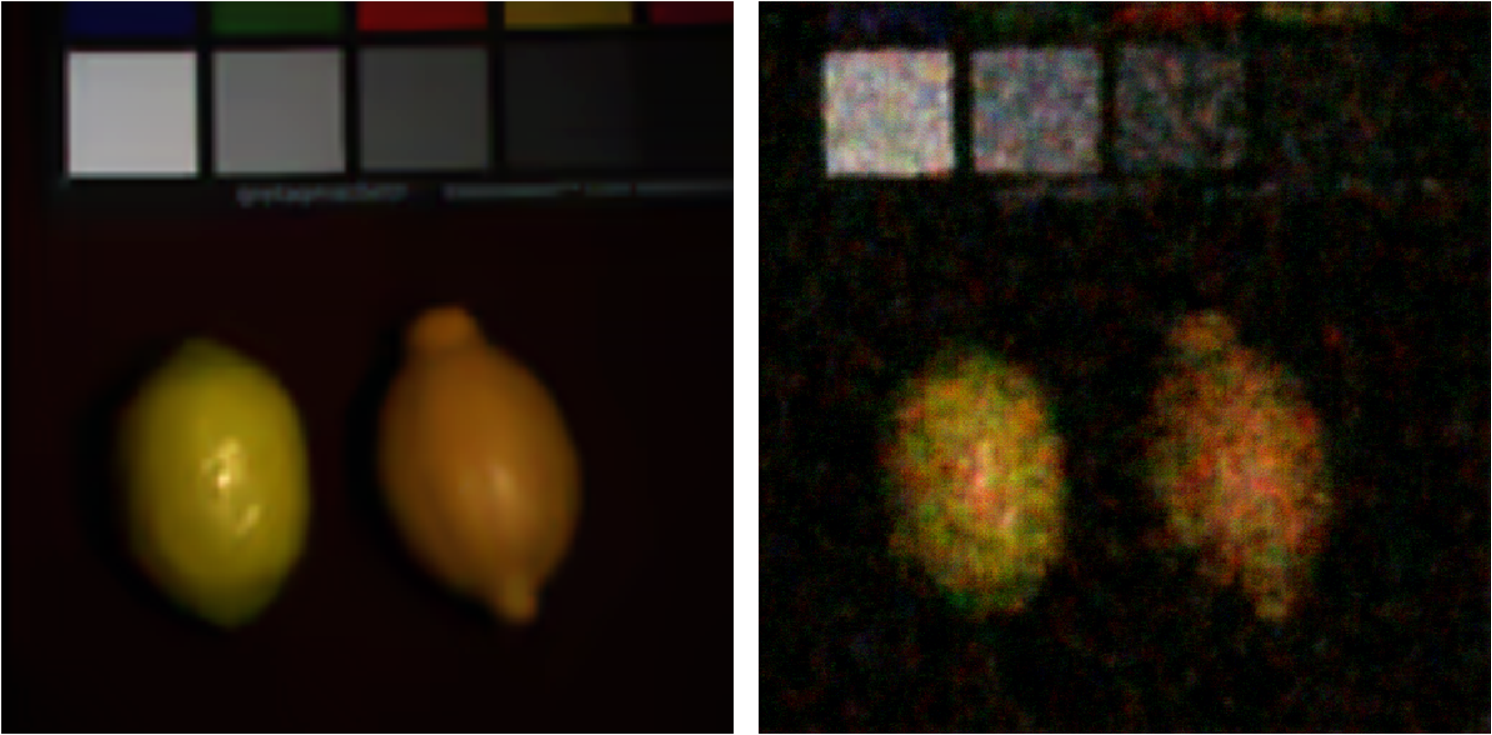}
    }
    \hfill
    \subfloat[Visualized result on DC Mall.]{\label{results_other} 
        \includegraphics[width=0.23\textwidth]{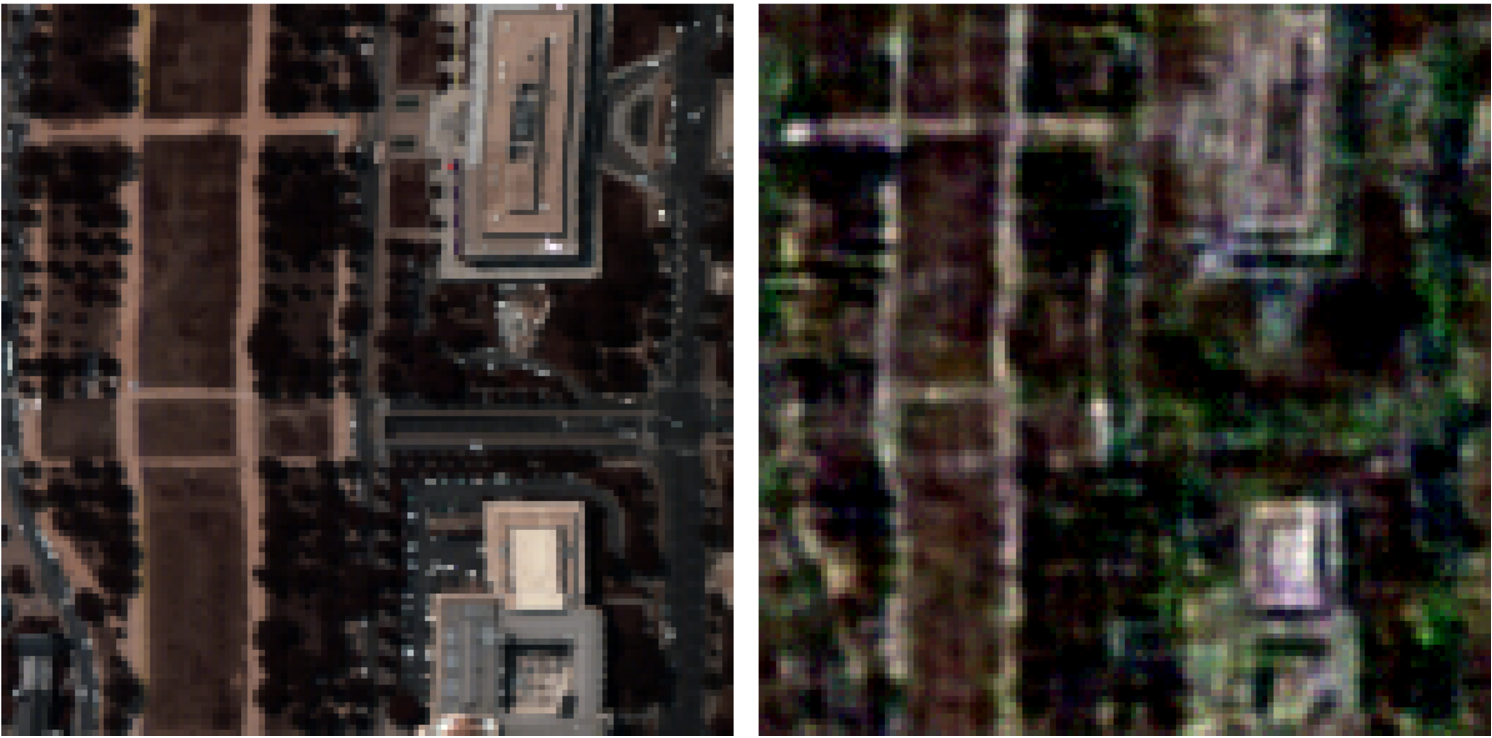}
    }
    \caption{Visualized experimental results on different datasets. The left image is RGB ground truth, while the right image is RGB from HSI Reconstruction.}
    \label{results_vis}
\end{figure}

As shown in Fig. \ref{results_1}, the experimental results demonstrate the PSNR performance of the proposed method compared to aligned-based frameworks \cite{10233481,10437917} and a conventional ``BPG+Capacity'' scheme under different channel conditions (AWGN, MIMO, and Rayleigh) for the CAVE and DC datasets. The proposed method consistently outperforms the aligned-based framework across all channel conditions. For the DC dataset, which contains richer spectral information, the proposed method achieves a higher PSNR than both the aligned-based framework and the ``BPG+Capacity'' scheme across all tested channels. In low-SNR conditions, the performance of the channel-adaptive mechanism may degrade due to the reduced quality of received features as shown in Fig. 5. To address this, integrating a diffusion-based model can serve as an effective strategy, particularly in low-SNR scenarios \cite{L2024Diffusion}. Experimental results demonstrate that our method outperforms the benchmark under all tested conditions.
% This approach iteratively removes noise from the received features, enhancing their quality and improving the mechanism's ability to adaptively integrate information under low-SNR conditions.

We also evaluate the performance across different $l$ values under varying SNR conditions. As illustrated in Fig. \ref{results_2}, the results show a clear improvement in both PSNR and SSIM when using the hierarchy-aware fusion approach compared to single-source methods. This enhancement underscores the effectiveness of the hierarchy-aware fusion approach in improving both the perceptual quality and accuracy of the reconstructed HR-HSI. Furthermore, visualized experimental results, as shown in Fig. \ref{results_vis}, display RGB images extracted from the reconstructed HSI.

\begin{table}[t]
\centering
\caption{The ablation study of the proposed data fusion semantic communication for different $l$, where the ``Basic Fusion'' has approximately 30M parameters, while the others have approximately 34M parameters.}
\begin{tabular}{|c|c|c|c|c|c|c|c|}
\hline
$l$  & Method & -3 dB & -1 dB & 1 dB  & 3 dB  & 5 dB  & 7 dB\\
\hline
\multirow{4}{*}{\!\!128\!\!}  
 & \!\!Full Fusion \!\!       & 19.7& 22.0& 23.6& 24.8& 25.5& 26.0 \\
 & \!\!Proposed \!\!          & 19.2& 21.2& 23.0& 24.2& 25.1& 25.6  \\
 & \!\!Seperate Fusion  \!\!     & 18.6& 20.8& 22.6& 23.8& 24.5& 25.0  \\
 & \!\!Basic Fusion \!\!      & 17.7& 19.8& 21.5& 22.9& 23.8& 24.4  \\
\hline
\multirow{4}{*}{\!\!32\!\!}   
 & \!\!Full Fusion \!\!       & 16.5& 18.9& 20.8& 22.4& 23.6& 24.4  \\
 &\!\! Proposed   \!\!        & 15.6& 18.0& 19.8& 21.5& 22.9& 24.0 \\
 & \!\!Seperate Fusion \!\!      & 14.7& 17.2& 19.2& 20.9& 22.4& 23.3 \\
 &\!\! Basic Fusion  \!\!     & 14.4& 16.5& 18.4& 20.0& 21.4& 22.3 \\
\hline
\end{tabular}
\label{ablation}
\end{table}
To evaluate the effectiveness of our proposed framework, we conducted an ablation study across SNRs ranging from $-3$ dB to $7$ dB, as shown in Table \ref{ablation}. The ``Full Fusion'' transmits all $\boldsymbol{S}_f$, $\boldsymbol{S}_s$, and $\boldsymbol{S}_p$ features without bandwidth constraints, while the ``Proposed'' employs the hierarchy-aware fusion module to integrate and transmits these features $\boldsymbol{S}$ and the vectors $\boldsymbol{M}_{\left(f,p\right)}$ and $\boldsymbol{M}_{\left(\left(f,p\right),s\right)}$. The ``Seperate Fusion'' method also uses the hierarchy-aware fusion module but does not send $\boldsymbol{M}_{\left(f,p\right)}$ and $\boldsymbol{M}_{\left(\left(f,p\right),s\right)}$ to the receiver, impacting reconstruction quality. Finally, ``Basic Fusion'' transmits only the basic fused features $\boldsymbol{S}_f$. The bandwidth usage for the ``Proposed'', ``Seperate Fusion'' and ``Basic Fusion'' methods is identical, while ``Full Fusion'' requires three times more bandwidth. Experimental results indicate that our method outperforms the ``Basic Fusion'', achieving an average PSNR increase of approximately $1.5$ dB. Compared to the ``Full Fusion'' method, our approach exhibits a marginal performance decrease of less than $0.5$ dB while reducing bandwidth consumption by two-thirds. This demonstrates the effectiveness of hierarchy-aware data fusion in ensuring high-quality HR-HSI reconstruction under bandwidth constraints. Specifically, the proposed method requires 486.49 GFLOPs with an overall latency of 0.11 seconds. The Transformer-based channel-adaptive mechanism contributes approximately 0.015 seconds of latency and an extra 31.36 GFLOPs in computational load. %Our proposed channel-adaptive fusion technique ensures reconstruction performance and significantly reduces bandwidth usage. We further enhance the reconstruction performance by integrating the proposed fusion strategies at the receiver, where the additional bandwidth costs are negligible.

%\begin{table} \normalsize
%\centering
%\caption{Comparison between the single source and the proposed data fusion semantic communication at the sizes of source data and the transmitted symbols and computational complexity, when $l=8$.}
%\begin{tabular}{c|c|c|c}
%\hline 
%\!\!\!\!Method\!\!\!\!&\!\!No. Symbols\!\!&\!\!FLOPs\!\!&\!\!Parameters\!\!\\
%\hline 
%\multirow{3}{*}{\!\!Single Source\!\!}
% & 31744 (LR-HSI) & \multirow{3}{*}{1.24T} & \multirow{3}{*}{13.88M} \\
% & 8126464 (HR-HSI) & &  \\
% & 131072 (Transmit) & & \\
%\hline
%\multirow{4}{*}{\!\!Data Fusion\!\!}
% & 31744 (LR-HSI) & \multirow{4}{*}{1.24T} & \multirow{4}{*}{13.89M} \\
% & 786432 (HR-RGB) & & \\
% & 8126464 (HR-HSI) & & \\
% & 131072 (Transmit) & & \\
%\hline
%\end{tabular}
%\label{tab:dimension}
%\end{table}
%Table \ref{tab:dimension} compares the single source and the proposed data fusion semantic communication in terms of the number of transmitted symbols and computational complexity when $l=8$. It can be seen that both methods have the same number of transmitted symbols. The computational complexity of the data fusion semantic communication is slightly higher than that of the single source semantic communication, but the difference is negligible. This is because the data fusion semantic communication contains some additional concatenate layers, which do not significantly increase the computational complexity.

\section{Conclusion}
In this letter, we proposed a novel method to address bandwidth challenges in hyperspectral super-resolution tasks based on the fusion of LR-HSI and HR-RGB within a semantic communication framework. Our approach integrates a hierarchy-aware and channel-adaptive data fusion mechanism, enhancing the transmission efficiency of fused information without increasing bandwidth consumption. The proposed hierarchy-aware fusion module adaptively combines deep and shallow features, ensuring that preserves structural and detailed information during reconstruction. Validated by significant improvements in PSNR and SSIM scores on the CAVE dataset, our approach demonstrated enhanced image quality and bandwidth efficiency. These results confirm the effectiveness of our semantic communication framework in enabling high-quality HR-HSI reconstruction while efficiently managing bandwidth.

\bibliographystyle{IEEEtran}
\bibliography{IEEEabrv,reference}

\end{document}